\pdfoutput=1
\documentclass[iop]{emulateapj}
\usepackage{epsfig}
\usepackage{graphicx}

\def\amm{NH$_3$}
\def\ammm{NH$_3$(3,3)}
\def\meth{CH$_3$OH}

\shorttitle{Ammonia (\amm) and Methanol (\meth) Emission Near Supernova
  Remnants} \shortauthors{McEwen et al.}

\begin{document}

\title{\ammm~and \meth~near Supernova Remnants: GBT and VLA
  Observations}

\author{Bridget C. McEwen, Ylva M. Pihlstr\"{o}m\altaffilmark{1}}
\affil{The Department of Physics and Astronomy, The University of New
  Mexico, Albuquerque, NM, 87131}

\altaffiltext{1}{Y.~M.~Pihlstr\"om is also an Adjunct Astronomer at the
 National Radio Astronomy Observatory}

\author{Lor\'{a}nt O. Sjouwerman} \affil{National Radio Astronomy
  Observatory, P.O. Box O, 1003 Lopezville Rd., Socorro, NM, 87801}

\begin{abstract}
  We report on Green Bank Telescope 23.87 GHz \ammm\, emission
  observations in five supernova remnants interacting with molecular
  clouds (G1.4$-$0.1, IC443, W44, W51C, and G5.7$-$0.0). The
  observations show a clumpy gas density distribution, and in most
  cases the narrow line widths of $\sim3-4$\,km\,s$^{-1}$ are
  suggestive of maser emission.  Very Large Array observations reveal
  36~GHz and/or 44~GHz \meth\, maser emission in a majority (72\%) of
  the \amm\, peak positions towards three of these SNRs. This good positional correlation is in
  agreement with the high densities required for the excitation of
  each line. Through these observations we have shown that \meth\,
  and \amm\, maser emission can be used as indicators of high density
  clumps of gas shocked by supernova remnants, and provide density
  estimates thereof. Modeling of the optical depth of the
  \ammm~emission is compared to that of \meth, constraining the
  densities of the clumps to a typical density of the order of
  $10^5$~cm$^{-3}$ for cospatial masers.  Regions of gas with this density are found to exist in the post-shocked gas quite close to the SNR shock front, and may be associated with sites where cosmic rays produce gamma-ray emission via neutral pion decay.
\end{abstract}


\section{Introduction}\label{intro}
Interactions between supernova remnants (SNRs) and molecular clouds
(MCs) can cause perturbations in the gas that may affect the evolution
of the surrounding interstellar medium. For example, it has long been
suggested that star formation might be triggered in the parent
cloud.  Details of such a proposed triggering processes have not been
clearly outlined and confirmed, partly due to the complexity of the
regions in the inner Galaxy where the SNR/MC interactions
are more probable.  Another plausible effect of SNR/MC collisions is the
acceleration of Galactic cosmic rays, since shock induced particle
acceleration can generate relativistic energies.  Models of cosmic ray
acceleration usually relate the brightness of resulting $\gamma$-ray
emission due to neutral pion decay using the gas
number density \citep[e.g.,][]{drury1994,abdo2010,cristofari2013}. The
density estimates deduced from $\gamma$-ray emission measurements are
often larger (10-20 times) than those inferred from X-ray
observations, indicating a strongly clumped medium is required to
produce the $\gamma$-ray emission \citep[e.g.,][]{slane2015}. Detailed
density estimates, and density gradients, in regions associated with
$\gamma$-ray emission are thus of interest to constrain the inputs of
cosmic ray acceleration models.  Observations of molecular transitions
of various molecules excited by the passing shock provides a tool to
derive densities, and can also provide kinematic information
of the gas. An overall understanding of the conditions in the
interacting cloud, both pre- and post-shock, may provide basic facts
to guide models of both SNR induced star formation as well as cosmic
ray acceleration.

\begin{deluxetable*}{llrrcrcc}]
\tabletypesize{\scriptsize}
\tablecaption{Observed SNRs \label{tbl-1}}
\tablewidth{0pt}
\tablehead{
\colhead{Source} &\colhead{Other Name}&\colhead{RA}& \colhead{DEC} &\colhead{SNR Size}& \colhead{V$_{LSR}$}&\colhead{Distance}&\colhead{GBT Channel noise}\\ 
&&\colhead{(J2000)} & \colhead{(J2000)}  &  \colhead{(RA\arcmin$\times$DEC\arcmin)} & \colhead{(km\,s$^{-1}$)}&\colhead{(kpc)}&\colhead{(mJy\,bm$^{-1}$)}
}
\startdata
G049.2$-$0.7 & W51C  & 19 23 50 & +14 06 00   & 30$\times$20 & 70.0   & 6.0 & 17\\
G005.7$-$0.0* &       & 17 59 02 & $-$24 04 00 & 8$\times$8   & 13.0   & 3.2 & 16\\  
G034.7$-$0.4 & W44   & 18 56 00 & +01 22 00   & 25$\times$30 & 45.0   & 2.5 & 16\\   
G189.1+3.0   & IC443 & 06 17 00 & +22 34 00   & 45$\times$45 & 5.0    & 1.5 & 12\\
G001.4$-$0.1 &       & 17 49 39 & $-$27 46 00 & 10$\times$10 & $-$2.0 & 8.5 & 18
\enddata
\label{snrinfo}
\tablenotetext{*}{SNR candidate \citep{Brogan2006}}
\end{deluxetable*}

\begin{figure*}[thb]
\centering
\includegraphics[width=1.03\textwidth]{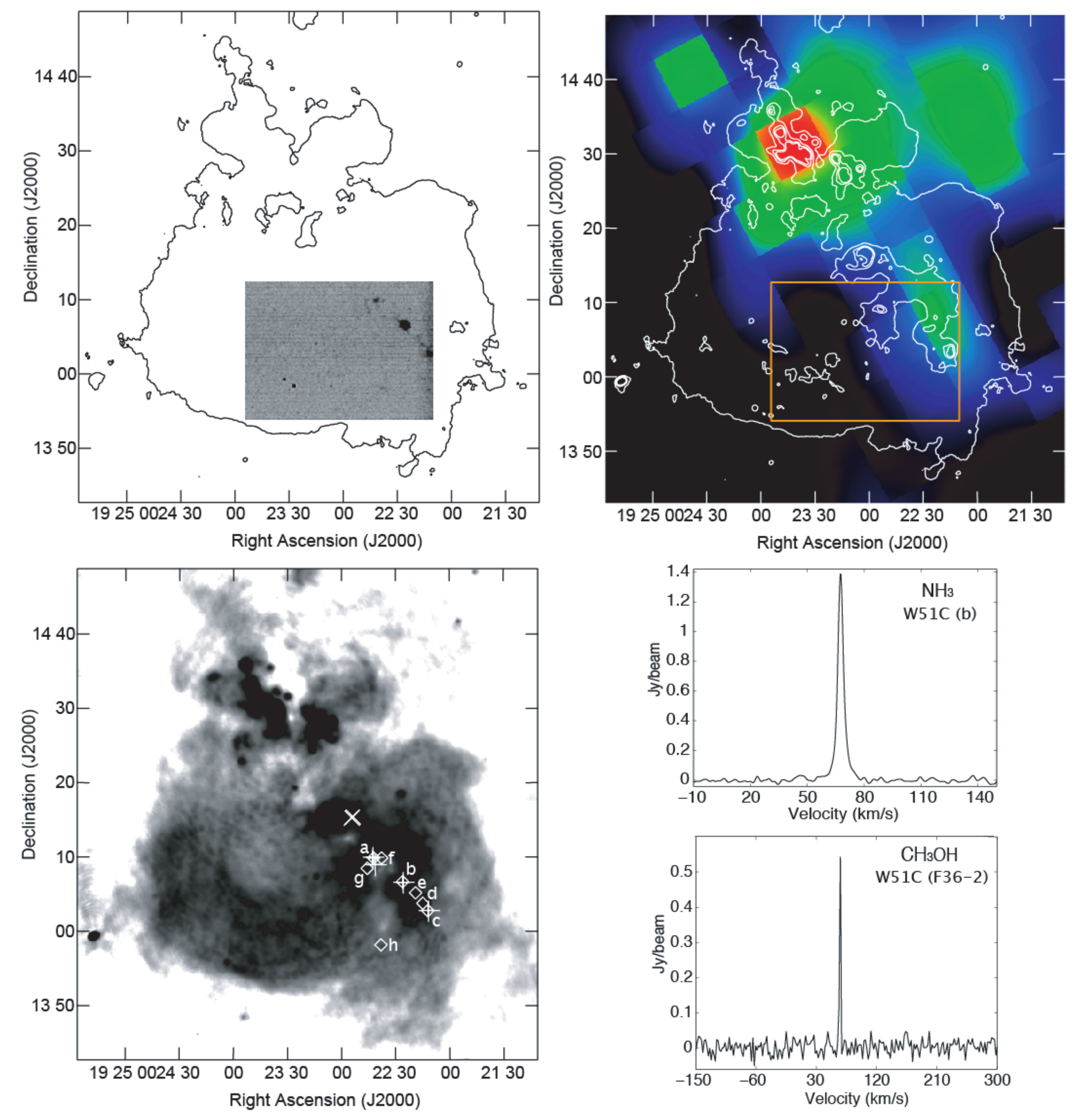}
\caption{{\it Top left}:  \amm~emission distribution (greyscale) outline by 90 cm radio continuum (contours at $30$ mJy\,beam$^{-1}$) of the W51 complex.  {\it Top right}:  $60-80$ km\,s$^{-1}$ CO(1$-$0) emission \citep{dame2001}, indicating the interacting MC, with respect to the radio continuum (levels at 30, 150, 300, and 400 mJy\,beam$^{-1}$).  {\it Bottom left}:  36 GHz CH$_3$OH masers (white plus signs) detected with the VLA with respect to \amm~emission peaks (white diamonds), 1720 MHz OH masers (white crosses), and continuum emission (greyscale).  {\it Bottom right}: Example spectral profiles of the brightest NH$_3$ emission region (b) and the brightest 36 GHz CH$_3$OH maser detected (F36-2).}
\label{w51call}
\end{figure*}

\begin{figure*}[thb]
\centering
\includegraphics[width=0.98\textwidth]{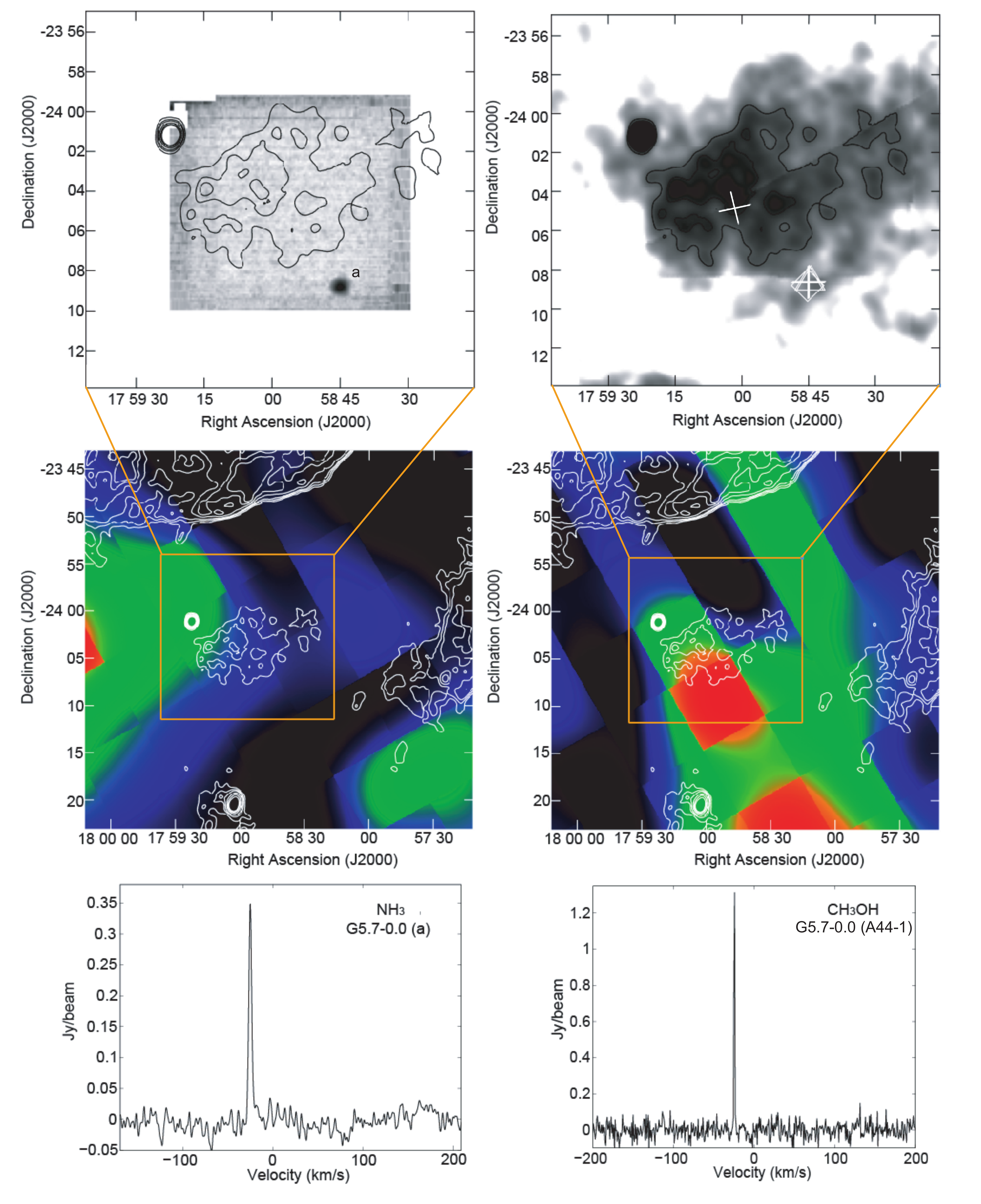}
\caption{{\it Top left}:  \amm~emission  distribution (greyscale) overlaid on a  90 cm radio continuum (contours at 25, 37.5, 50, 75, 100, and 125 mJy\,beam$^{-1}$) of G5.7$-$0.0. {\it Top right}:  36 GHz (white plus sings) and 44 GHz (white triangles) CH$_3$OH masers  detected with the VLA with respect to \amm~emission peaks (white diamonds), 1720 MHz OH masers (white crosses), and continuum emission (greyscale).  {\it Middle left}: $3-23$ km\,s$^{-1}$ CO(1$-$0) emission \citep{dame2001}, which indicates the location of an interacting MC around the velocity of the OH maser ($\sim13$ km\,s$^{-1}$), with respect to the radio continuum (contours). {\it Middle right}:  $-16$ to $-36$ km\,s$^{-1}$ CO emission, which indicates the locations of another interacting MC around the velocities of the NH$_3$ and CH$_3$OH ($\sim-26$ km\,s$^{-1}$), with respect to the radio continuum (contours). {\it Bottom left}: Example spectral profiles of the NH$_3$ emission region (a).  {\it Bottom right}: Example spectral profiles of the brightest 44 GHz CH$_3$OH maser detected (A44-1).}
\label{g57all}
\end{figure*}

\begin{figure*}[thb]
\centering
\includegraphics[width=1.03\textwidth]{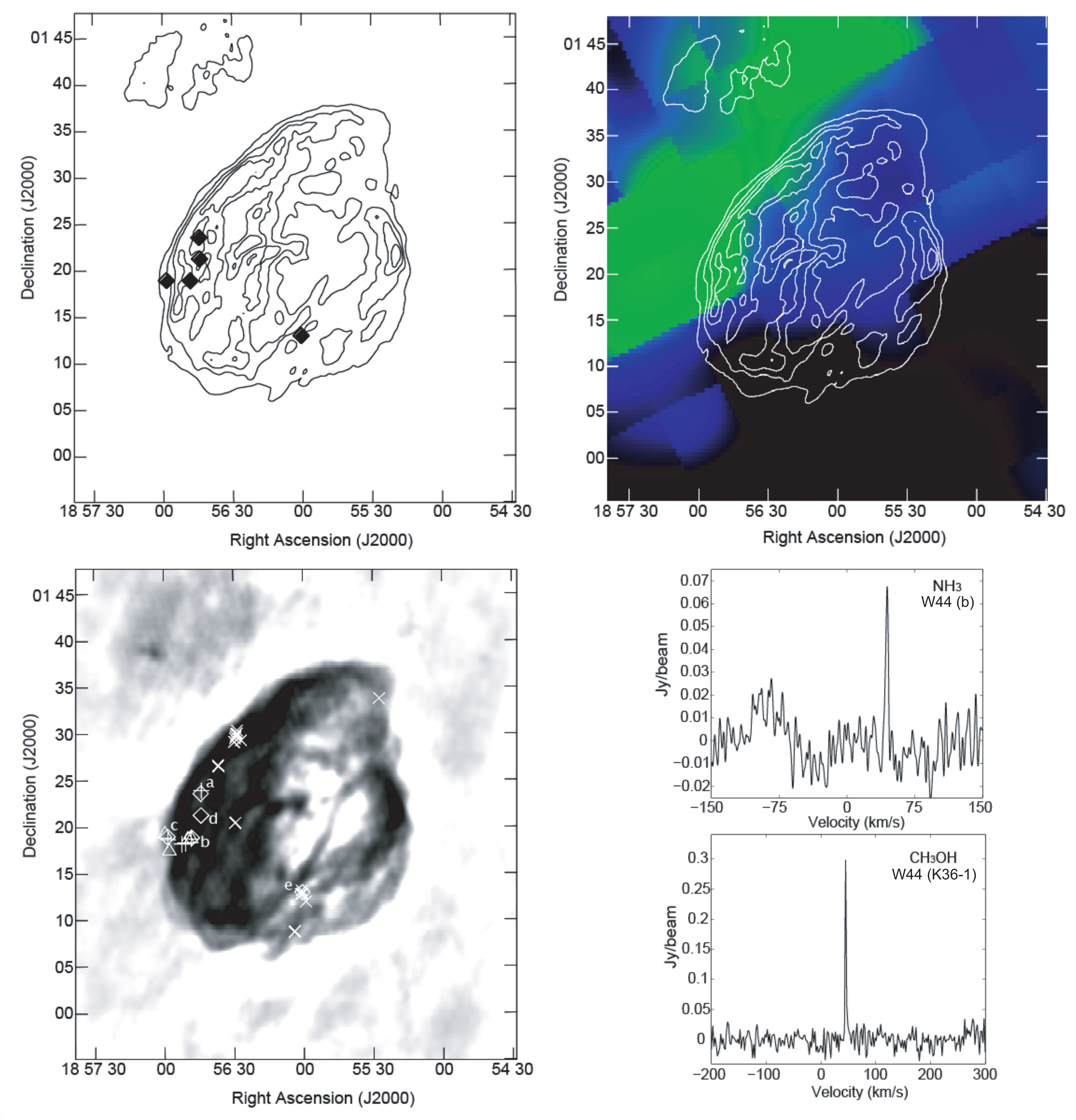}
\caption{{\it Top left}:  \amm~emission peaks (black diamonds) overlaid on 21 cm radio continuum (contours at 5, 25, 50, 100 and mJy\,beam$^{-1}$) of  W44. {\it Top right}:  $35-55$ km\,s$^{-1}$ CO emission \citep{dame2001}, which indicates the location of the interacting MC, with respect to the radio continuum (contours). {\it Bottom left}:  36 GHz (white plus signs) and 44 GHz  (white triangles) \meth~emission with respect to the \amm~emission peaks (white diamonds), 1720 MHz OH masers (white crosses), and continuum emission (greyscale).   {\it Bottom right}: Example spectral profiles of the brightest NH$_3$ emission region (b) and the brightest 36 GHz CH$_3$OH maser detected (K36-1).}
\label{w44all}
\end{figure*}

\begin{figure*}[thb]
\centering
\includegraphics[width=1.03\textwidth]{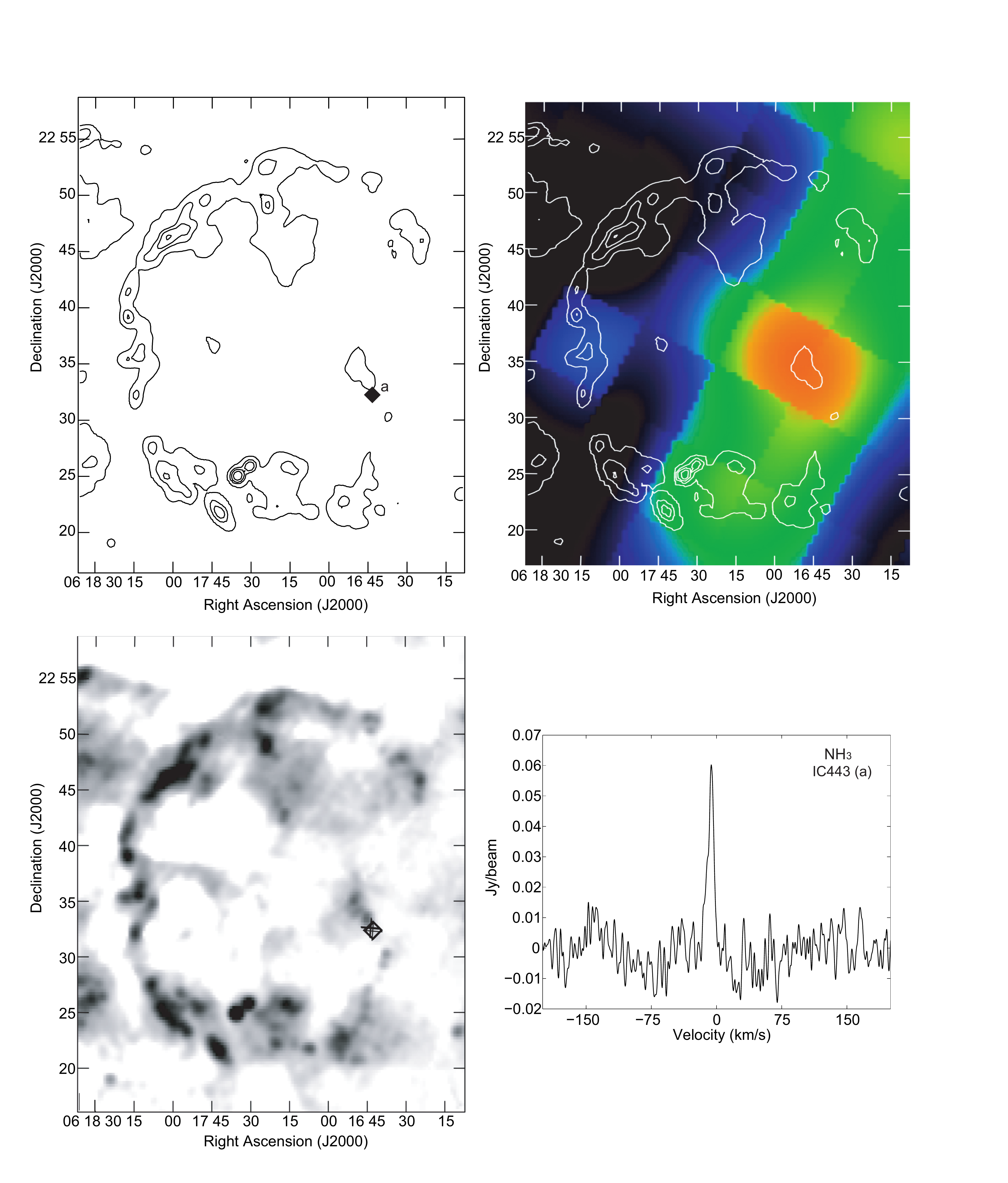}
\caption{{\it Top left}:   \amm~emission peak (black diamond) overlaid on 21 cm radio continuum (contours at 20, 80,140, and 200 mJy\,beam$^{-1}$) of  IC443. {\it Bottom left}:  \amm~emission peak (black diamond) and 1720 MHz OH masers (black plus signs) with respect to the continuum emission (greyscale).  {\it Top right}:  $-5$ to 15 km\,s$^{-1}$ CO emission \citep{dame2001}, which indicates the location of the interacting MC, with respect to the radio continuum (contours). {\it Bottom right}: Example spectral profile of the NH$_3$ emission region (a).}
\label{ic443all}
\end{figure*}

\begin{figure*}[thb]
\centering
\includegraphics[width=0.95\textwidth]{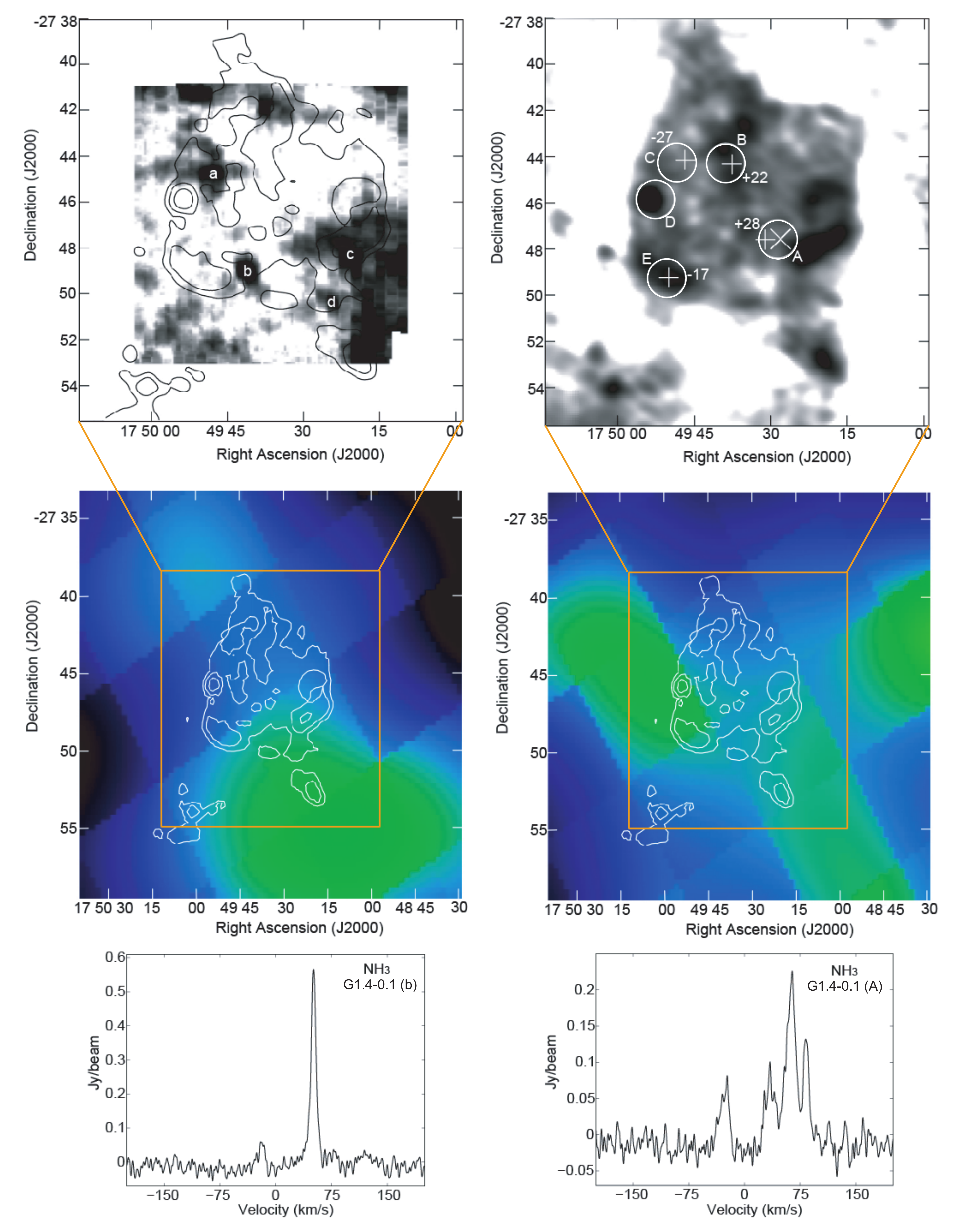}
\caption{{\it Top left}:  \amm~emission distribution (greyscale) outline by 21 cm radio continuum (contours at 4.3, 8.6, 17.2, and 34.4 mJy\,beam$^{-1}$) of G1.4$-$0.1.  The bright NH$_3$ regions (labeled {\it a} through {\it d}) have spectral features listed in Table 2.  {\it Top right}:  Average positions of the 36 GHz CH$_3$OH masers (white plus sings) in each pointing (white circles labeled with the letters {\it A} through {\it E} and the average velocity in km\,s$^{-1}$) detected by \citet{Pihl2014} with respect to 1720 MHz OH masers (white crosses), and continuum emission (greyscale).  {\it Middle left}: $20-30$ km\,s$^{-1}$ CO emission \citep{dame2001}, which indicates the location of an interacting MC, with respect to the radio continuum (contours). {\it Middle right}:  $-15$ to $-30$ km\,s$^{-1}$ CO emission, which indicates the locations of another interacting MC, with respect to the radio continuum (contours). {\it Bottom row}: Example spectral profiles of the NH$_3$ emission in region b (left) and region A (right).}
\label{g14all}
\end{figure*}

SNR/MC interactions are identified in different ways, for example, via shock excited OH 1720 MHz emission \citep{Claussen1997,Wardle2002}, near-infrared H$_2$ emission, and broad molecular line
widths exceeding those of cold molecular gas \citep{Reach2005}. In addition to the 1720 MHz OH, other shock excited masers detected in SNR/MC regions include
the Class I 36~GHz and 44~GHz methanol (\meth) masers, which
can be used to estimate densities in the maser emitting region
\citep{Sjou2010, Pihl2011, Pihl2014, Mcewen2014}. In characterizing
physical parameters of gas clouds in general, the ammonia (\amm)
molecule is commonly used as a probe of dense ($n\geq10^3$ cm$^{-3}$)
environments. Ratios of the line intensities allow estimates of
rotational excitation temperatures, heating, and column densities
\citep[e.g.,][]{Morris1983, Ho1983, Okumura1989,
  Coil1999}. Illustrating examples of where \amm~information can be
combined with that of maser species include the high resolution
observations of \amm~in the inner 15 pc of the Galactic center (GC)
region \citep{Coil1999, Coil2000, McGary2001}, and the recent
detections of Class I 36~GHz and 44~GHz \meth~maser emission toward
Sgr\,A\,East \citep{Sjou2010, Pihl2011, Mcewen2016}. These
observations reveal a close positional agreement between the
\amm~emission peaks and the location of \meth~masers. A similar
spatial coincidence between \amm~and 44~GHz \meth~masers has also been
found toward W28 \citep{Sjou2010, Nicholas2011, Pihl2014}.

A relatively general picture of the distribution of OH and \meth~
masers in SNRs comes from observations of Sgr\,A\,East, W28, and
G1.4$-$0.1 \citep[e.g.,][]{Sjou2010, Pihl2011, Pihl2014, Mcewen2014}.
Based on these observations, it has been shown that CH$_3$OH masers
are typically found offset from 1720 MHz OH masers \citep[e.g.,][]{Claussen1997, Frail1998, Yusef2003, Sjou2010, Pihl2014}.
This indicates that particular maser species and transitions are
tracing different shocked regions, or alternatively, different regions
in one shock. Consistently, bright 36~GHz \meth~masers have been found
to trace regions closer to the alleged shock front and OH masers are
more likely to trace the post-shocked gas. The location of the 44~GHz
masers relative to the shock front has been less clear.

To further develop the general picture of the pre- and post-shock gas
structures in SNR/MCs, including the presence and spatial positions of
different molecular transitions through the interaction region, the
NRAO Green Bank Telescope (GBT) was used to survey five SNRs (W51C, W44, IC443, G1.4$-$0.1, and G5.7$-$0.0) for
\amm~emission (Sect.~\ref{gbt}). Here we report on the spatial
distribution of the \ammm~emission compared to that of the 36 and
44~GHz \meth~maser emission previously observed by \citet{Pihl2014} in G1.4$-$0.1
and the emission detected in new observations in W51C, W44, and G5.7$-$0.1 using the Very Large
Array, VLA (Sect.~\ref{vla}).  Section \ref{results} discusses the results from the observations.  Estimates of the temperature and density
ranges most suitable for the formation of \ammm~masers in an SNR
environment are provided using model calculations (Sect.~\ref{modeling}),
and compared to those of the \meth~masers (Sect.~\ref{discussion}).

\section{Observations and Data Reduction}

\subsection{GBT Observations and Data Reduction}\label{gbt}
Observations were carried out using the GBT in January through May of
2013 under the project code GSLT051174. An original list of 17
possible targets was considered based on previously known
SNR/MC interaction sites (detected via 1720 MHz OH
masers). The allocated time allowed four targets to be fully mapped
(G5.7$-$0.0, IC443, W44, and G1.4$-$0.1) and one target to be
partially mapped (W51C).  Characteristics of each SNR observed, such
as size and systemic velocity, can be found in Table \ref{tbl-1}. It should be noted that the source G5.7$-$0.0 has not yet been confidently classified as an SNR because it is very faint in the radio, and does not have a typical shell-like morphology  \citep{Brogan2006}.  It is considered a SNR candidate since it displays non-thermal emission and is associated with an 1720 MHz OH maser, indicative of a MC/SNR shocked interaction \citep{Hewitt2009}. The sources
were surveyed for \amm~emission using the 7-beam K-band focal plane
array (KFPA) and spectrometer. The observations were made with dual
polarization, 50~MHz of bandwidth having a total velocity coverage of
about 600~km\,s$^{-1}$ and a velocity resolution of about 0.5 km
s$^{-1}$. Venus, Jupiter, Saturn, and the Moon were used as
calibrators depending on which source was visible at the time of each
observation. The DECLAT and RALONG standard spectral line survey
mapping modes were used, with a slew rate of around
3.6\arcsec\,s$^{-1}$.  In-band and out-of-band frequency switching
were utilized depending on the proximity of the SNR to the GC,
assuming possible velocities $\pm 200$ km\,s$^{-1}$, or more for
$|\ell|< 25^{\circ}$.

The data reduction and calibration was carried out using the GBT
pipeline following standard procedures outlined in the GBTPIPELINE
User's Guide. The GBT pipeline procedure was used to automate the
calibration and combine the data from multiple observing sessions. The
pipeline estimates the atmospheric opacity by using real-time weather
monitored data. Astronomical calibrator sources were used to correct
the system temperatures of the beams by applying gain factors (one for
each beam and polarization) that were acquired using {\it venuscal},
{\it jupitercal}, {\it saturncal}, and {\it mooncal} procedures in
GBTIDL.  In a few instances the reported system temperatures where
corrupted, in which case average gain factors were calculated from all
observing sessions combined. This caused a slightly larger error of
the absolute flux density value measured, but this does not affect the
results reported on in this paper. Finally, the beginning and ending
channels were clipped during the pipeline procedure, which varied for
each data set.

Channel averaging, continuum subtraction, and imaging were carried out
in AIPS using standard data reduction and imaging tasks applied to
single dish data. Every two channels were averaged and all the data
were Hanning-smoothed.  A third-order polynomial fit was subtracted
from the line free channels over the entire bandwidth.  An additional third-order polynomial fit was subtracted (outside AIPS) from the NH$_3$ emission in W44 and IC443 to make the baselines more linear, which we suspect were affected by severe weather conditions at the time of the observations.  Some of the
SNRs are very large and the full mapping could not be completed during one observing
session, therefore, the mapped regions were produced by combining
different data sets from different observing sessions. As a result,
the root mean square (RMS) noise level varies across these maps, which
can primarily be attributed to the varying weather conditions on the
different days of observation.  The average RMS noise values in a line
free channel for each target using the GBT are listed in column 8 of
Table \ref{tbl-1}.

\begin{deluxetable*}{lrrrrrccc}[thb]
\tabletypesize{\scriptsize}
\tablecaption{NH$_3$ Emission Information \label{tbl-2}}
\tablewidth{0pt}
\tablehead{
\colhead{NH$_3$ pos} &\colhead{RA}& \colhead{DEC} &\colhead{V$_{p}$}& \colhead{$\Delta$V}&\colhead{I$_{p}$}&\colhead{CH$_3$OH Maser}&\colhead{VLA }&\colhead{Comment}\\ 
&\colhead{(J2000)} & \colhead{(J2000)} & \colhead{{(km s$^{-1}$})}&\colhead{{(km s$^{-1}$})} &\colhead{{(Jy bm$^{-1}$})}&\colhead{Association (GHz)}&\colhead{Data}}&
\startdata

W51Ca  & 19 22 42.15 &  $+$14 09 54.0 &   60.7 & 4.1 & 0.12&36&Y&\\ % %%%%%most left bright Gaussian Done*
W51Cb  & 19 22 26.07 & $+$14 06 42.0&   67.5 & 3.4 & 1.39&36 \& 44&Y&Fig.\,1\\%%%had to eye ball it, gaussian did work %%middle bright
W51Cc  & 19 22 12.07 &  $+$14 02 48.0 &   68.7 & 3.4 & 0.19&36 \& 44&Y&\\   %%%most right *
W51Cd  & 19 22 15.36  & $+$14 03 48.0 &   71.2 & 4.3 & 0.10&44&Y&\\   %%2nd right*
W51Ce  & 19 22 19.07 &   $+$14 05 12.0 &   70.8 & 4.0 & 0.07&44&Y&\\ %3rd right*
W51Cf  & 19 22 37.62  &  $+$14 09 54.0&   66.9 & 3.1& 0.06&$^{**}$&Y&\\ %% to the right of the brightest left one*
W51Cg  & 19 22 46.28 &  $+$14 08 18.0 &   66.9 & 3.1 & 0.06&$^{**}$&Y&\\ %most left clump*
W51Ch  & 19 22 38.05 & $+$13 58 12.0 &   61.0 & 4.0& 0.06&$^{**}$&Y&\\%% lowest middle*
\noalign{\smallskip}\hline\noalign{\smallskip}
G5.7-0.0a&17 58 45.05 &  $-$24 08 48.0& $-$25.1& 3.7 & 0.37&36 \& 44&Y&Fig.\,2\\  %%%GAUSSIAN FIT DONE!!!!!!!!!!!!*
\noalign{\smallskip}\hline\noalign{\smallskip}
W44a   & 18 56 44.21  & $+$01 23 34.7 &   40.9 & 4.4 & 0.06& 36&Y&\\  %%%most top  %%%%%%had to eyeball it*
W44b   & 18 56 48.21 &$+$01 18 52.7 &   44.6 & 5.5 & 0.07&36 \& 44&Y&Fig.\,3\\ %%lower right*
W44c   & 18 56  58.62  & $+$01 18 46.7&   46.5 & 4.0 & 0.05&36 \& 44&Y&\\%%%lower left*
W44d   & 18 56 46.21 & $+$01 21 16.7 &   41.9 & 3.4 & 0.05&$^{**}$&Y&\\%%%middle right (small)*
W44e   & 18 56 01.80 & $+$01 12 40.7 &   44.0 & 3.6 & 0.06&$^{*}$&Y&\\%%%bottom middle lower of pic*
\noalign{\smallskip}\hline\noalign{\smallskip}
IC443a & 17 16 42.97 &  $+$22 32 23.9 & $-$6.1 & 6.7 & 0.06&$^{*}$&Y&Fig.\,4\\ %(beam area 30.42 pixles) *
\noalign{\smallskip}\hline\noalign{\smallskip}
G1.4-0.1a &17 49 48.14 &  $-$27 44 42.0& $-$31.7&11.7 &0.16&36&Y&\\%G1.4 NH3 clump a (labeled in the fig with 'a'
&&&47.4&3.7&0.48&36&&\\
G1.4$-$0.1b&17 49 41.36  &  $-$27 48 60.0&$-$20.6&8.3&0.07&$^{***}$&N&Fig.\,5\\%G1.4 NH3 clump b (labeled in the fig with 'b'
&&&50.5&7.4&0.58&$^{***}$&&\\
G1.4$-$0.1c&17 49 20.10&   $-$27 48 24.0&$-$37.5&16.0&0.17&$^{***}$&N&\\%G1.4 NH3 clump c (labeled in the fig with 'c'
&&&69.8&15.3&0.48&$^{***}$&&\\
G1.4$-$0.1d& 17 49 24.62 &  $-$27 50 18.0&$-$9.7&12.6&0.09&$^{***}$&N\\%G1.4 NH3 clump d (labeled in the fig with 'd'
&&&25.4&10.4&0.40&$^{***}$&&\\
&&&87.9&13.2&0.25&$^{***}$&&\\
G1.4$-$0.1A&17 49 31.09&$-$27 47 36.3&$-$23.4&11.7&0.10&36&Y&Fig.\,5\\%NH3 info at the position of the CH3OH maser in pointing A
&&&34.6&13.3&0.11&36&&\\
&&&64.3&13.2&0.24&36&&\\
&&&82.7&8.9&0.15&36&&\\
G1.4$-$0.1B&17 49 37.59& $-$27 44 20.4&$-$31.7&13.2&0.61&36&Y&\\%NH3 info at the position of the average CH3OH maser in pointing C
G1.4$-$0.1C&17 49 46.79& $-$27 44 08.8&$-$31.0&15.9&0.12&36&Y&\\%NH3 info to the position of the average CH3OH maser in pointing B
&&&44.7&10.4&0.27&36&&\\
G1.4$-$0.1E&17 49 49.97& $-$27 49 14.9&$-$18.5&14.1&0.12& 36&Y&\\%NH3 info to the position of the average CH3OH maser in pointing E
\noalign{\smallskip}\hline\noalign{\smallskip}
\noalign{$^{*}\phantom{^{**}}$ Searched in a previous study by \citet{Pihl2014} but  no CH$_3$OH was detected.}
\noalign{$^{**}\phantom{^{*}}$ Searched in this study but no CH$_3$OH was detected.}
\noalign{$^{***}$ This area has not been searched to date.}
\enddata
\label{snrnh3}
\end{deluxetable*}

\subsection{VLA Observations and Calibration}\label{vla}
The VLA was used to search for \meth~emission towards the \amm~peak positions found with the
GBT (excluding positions previously searched).  Note that one of the primary reasons for the GBT observations was to obtain positions of NH$_3$ peaks.   Even though some of the detected NH$_3$ emission was very weak, e.g., in W44, they were still included in the VLA search.  The data were acquired on September 4, 2014 (project code
14A-191) using the Ka$-$band and Q$-$band receivers. To cover
the spatial extent of the \amm~emission, 1(1), 11(17) and 12(14)
pointings at 36(44) GHz were used in G5.7$-$0.0, W44 and W51C,
respectively. The array was in D configuration resulting in typical
synthesized beam sizes of 2.65\arcsec $\times$1.90\arcsec~at 36~GHz
and 2.11\arcsec $\times$1.84\arcsec~at 44~GHz.  The VLA primary beam
is 1.25\arcmin~at 36~GHz and 1.02\arcmin~at 44~GHz.

512 frequency channels were used across a 128~MHz bandwidth (channel
separation close to 1.9 km\,s$^{-1}$), centered on the sky frequency
of the given target estimated using the previously known OH maser
emission.  The data were calibrated and imaged using standard
procedures in AIPS pertaining to spectral line data. For each pointing
position, the on-source integration time was approximately 1(2)
minutes at 36(44) GHz. The precise integration time varied somewhat
between the positions depending on how the observations fit within the
scheduling block. The typical final RMS noise values were between
about 12 and 40~mJy per channel, depending on the observed frequency
and observing conditions. Under ideal weather conditions, the theoretical RMS noise would be close to 10 mJy\,beam$^{-1}$.  Degraded weather and a varying number of antennas in the array is consistent with the slightly higher measured RMS noise.  No 36~GHz or 44~GHz continuum emission was
found in any of the observed fields.

\begin{deluxetable*}{lrrrrrrcl}[thb]
\tabletypesize{\scriptsize}
\tablecaption{CH$_3$OH Maser Emission Information \label{tbl-3}}
\tablewidth{0pt}
\tablehead{
\colhead{Source}&\colhead{$\nu$} &Name &\colhead{RA}& \colhead{DEC} &\colhead{V$_{p}$}& \colhead{$\Delta$V}&\colhead{I$_{p}$}\\ 
&\colhead{(GHz)}&&\colhead{(J2000)} & \colhead{(J2000)} & \colhead{({km s$^{-1}$})}&\colhead{({km s$^{-1}$})} &\colhead{({Jy bm$^{-1}$})}
}
\startdata
W51C&36&B36-1&  19 22 42.69& $+$14 09 55.5 &57.6&6.0&0.25\\%
&36 &B36-2& 19 22 42.21  &$+$14 09 51.5   & 59.6 & 4.7 &0.14\\%
& 36& B36-3& 19 22 42.76 & $+$14 10 02.0 &57.6 &3.0&0.13\\%
&36&  B36-4& 19 22 41.42 &  $+$14 09 03.5 &74.1&10.0&0.24\\%
&36& F36-1& 19 22 25.70 & $+$14 06 35.0 &65.9 &3.1& 0.43\\%
&36& F36-2& 19 22 26.01 &  $+$14 06 36.0&65.9&2.5&0.54\\%
&36& F36-3& 19 22 26.11 & $+$14 06 37.0 &63.8&4.0&0.50\\%
&36& L36-1 & 19 22 12.08 & $+$14 02 49.5 &67.9&4.0&0.43\\
&36& L36-2 & 19 22 07.34 & $+$14 02 18.5 &76.2&4.0&0.19\\
\noalign{\smallskip}\hline\noalign{\smallskip}
G5.7$-$0.0&36&A36-1&17 58 44.99 & $-$24 08 38.0 & $-$26.4 & 3.5& 0.38\\
 & 36 &A36-2& 17 58 45.17 &  $-$24 08 39.5 & $-$26.4 & 4.0 & 0.26\\
 & 36 &A36-3 & 17 58 44.85  & $-$24 08 36.5 & $-$26.4 & 4.0 & 0.21\\
&44&A44-1& 17 58 44.48&  $-$24 08 37.0& $-$24.4&1.8&1.31\\
&44 & A44-2& 17 58 44.88 &  $-$24 08 37.0 &$-$24.4&3.0&0.35\\
\noalign{\smallskip}\hline\noalign{\smallskip}
W44&36 &A36-1& 18 56 43.96 &  $+$01 23 54.2 & 40.9 & 3.0&0.25\\%
&36 &F36-1&18 56 48.39 &  $+$01 18 45.7 & 42.9 &2.0&0.09\\%
&36& G36-1& 18 56 50.82 &  $+$01 18 17.3& 42.9&4.5&0.14\\%
&36 & G36-2&18 56 50.86 &  $+$01 18 21.8 & 40.9&4.5&0.11\\%
&36 & G36-3& 18 56 52.49 & $+$01 18 12.3 & 47.1 &7.0 & 0.12\\%
&36& K36-1&18 56 58.46 &  $+$01 18 41.7 &45.0 &2.7 &0.30\\%
&44& K44-1&18 56 49.86 &   $+$01 18 43.2 &56.9& 0.7 &0.21\\%
&44& L44-1& 18 56 48.36 &   $+$01 18 46.2 &45.0 &1.6 &0.31\\%
&44& P44-1&18 56 57.86 &    $+$01 17 27.7 & 35.6& 1.1& 0.26\\%
&44& Q44-1 & 18 56 59.69 &    $+$01 19 17.2 &6.7 &1.5&0.24%

\enddata
\label{snrch3oh}
\end{deluxetable*}

\subsection{Radio Continuum and CO Line Images}
To visualize the spatial distribution emission across each SNR,
low-frequency radio continuum maps were assembled from archival
data. The 21 cm IC443 map was taken from the NRAO VLA
Sky Survey (NVSS; Condon et al. 1998). The 21~cm images of W44 and
G1.4$-$0.1 were made from VLA archive data taken in 1984 and 1995,
respectively, using standard AIPS calibration and imaging
procedures. The 90cm W51C and G5.7$-$0.0 images were obtained and adapted from
\citet{Brogan2013} and \cite{Brogan2006}, respectively.  A comparison of the NH$_3$ and CH$_3$OH distribution with CO(1$-$0) is also presented, using images from the \citet{dame2001} Milky Way CO survey.

\section{Results}
\label{results}

\subsection{\amm~Emission}\label{gbtres}
\ammm~emission was detected towards all the SNRs observed.  Four of
the targets (G5.7$-$0.0, W44, W51C, and IC443) display compact clumps
of emission, with the majority having relatively narrow line widths of
$\sim 3-4$~km\,s$^{-1}$, see Figs.~\ref{w51call}-\ref{ic443all}. G1.4$-$0.1 has a much more extended
distribution of the emission, and has peaks combined into broader
lines with widths closer to $8-16$~km\,s$^{-1}$ (Fig.~\ref{g14all}). The peak position,
velocity centroid, velocity width, and the peak flux density of each
emission region is listed in Table \ref{tbl-2}. For each SNR, the
brightest spectral profiles are plotted in Figs.~\ref{w51call}-\ref{g14all}.  Included in Figs.~\ref{w51call}-\ref{g14all} are the spatial positions of the NH$_3$ emission regions overlaid on radio continuum maps outlining details and boundaries of each  SNR (in greyscale and contours).  In addition, the radio continuum is overlaid on CO(1$-$0) maps averaged over the velocities corresponding to those of the NH$_3$ emission.  This shows the interacting MCs with respect to each SNR.

The narrow linewidths may indicate maser emission from the metastable
\ammm~transition. In the case of G1.4$-$0.1, the extended emission
combined with broader linewidths may be the result of maser emission
mixed with thermal emission. The angular resolution of the GBT is not
sufficient to set strong limits of the brightness temperature, which
could have provided further information on the possibility of maser
emission. With no data from the lower metastable states (the (1,1) and
(2,2) transitions) and with limited resolution, further studies to
investigate the candidate maser will have to be performed.

\subsection{\meth~ Maser Candidates} \label{vlares} 

\meth~maser emission was detected in two SNRs (W44 and W51C) and in the one candidate SNR (G5.7$-$0.0) observed. Typical FWHMs range from about
0.7 to 10~km\,s$^{-1}$. Features labeled as detections all had a
signal-to-noise ratio greater than 5 in the centroid channel, and
emission was detected in three or more channels.  The positions, centroid velocities, FWHMs, and peak flux densities of the candidate maser features found in our new VLA observations are listed in Table \ref{tbl-3}. The FWHMs are narrow,
and example spectral profiles of the brightest maser associated with each SNR can be found in Fig.~\ref{w51call} to \ref{w44all}.  Note that due to the combination of a small VLA field-of-view and the limited observing time, the CH$_3$OH emission search did not cover the entire SNR/MC interaction regions.

The emission regions
were unresolved with the VLA, except for the 36~GHz emission in
G5.7$-$0.0.  The emission in this source was marginally resolved and
had three peak flux density positions. \meth~maser candidates were
detected in the majority of the pointings. Although 44~GHz maser
candidates were visible towards W51C, we chose not to include those in
Table \ref{tbl-3}, due to a significantly increased noise in this data
set caused by a standing wave feature of unknown origin.

\subsection{\meth~and \amm~Association}
Table \ref{tbl-2} lists the \amm~regions that have positional
association with 36 and/or 44 GHz \meth~candidate masers. The
brightest \amm~regions associated with W44, W51C, and G5.7$-$0.0 have
both 36 and 44 GHz \meth~candidate maser association. Overall, 14 out
of 20 (72\%) positions observed in both the \ammm~and \meth~display
emission from both molecular species.

\section{Optical Depth Modeling}\label{modeling}
\begin{figure}[tbh]
\center
\includegraphics[width=9cm]{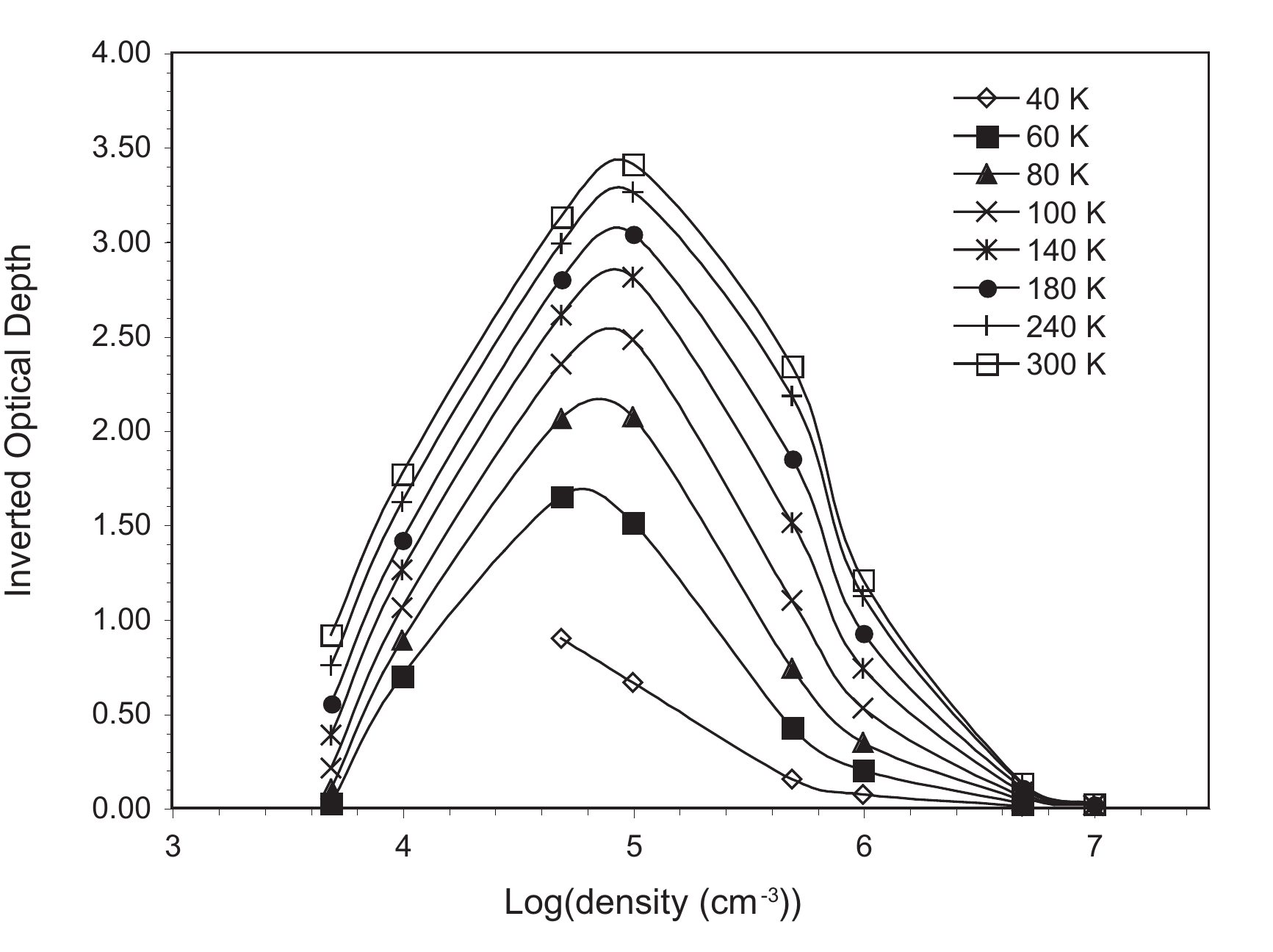}
\caption{Peak inverted optical depth versus number density of H$_2$,
  with a clear peak for densities around 10$^5$ cm$^{-3}$.}
\label{model}
\end{figure}

The relative intensity of co-located molecular emission transitions
can be used to estimate the physical condition of the gas. In
\citet{Mcewen2014}, the MOLPOP-CEP\footnote{Available at
  http://www.iac.es/proyecto/magnetism/pages /codes/molpop-cep.php} code for coupled radiative transfer and level
population calculations was used to estimate the conditions conducive for 36~GHz and 44~GHz \meth~masers in SNRs. To compare the range
of temperature and densitites producing \ammm~maser emission with that
of the \meth~maser, the same code was used to determine the optical
depth of the \ammm~transition.  Following the same assumptions of the SNR
conditions as in \citet{Mcewen2014}, an external radiation field of
2.725 K from the CMB and a 30 K dust radiation field were used
in the models. Energy levels for the \ammm~molecule were
incorporated using the Leiden Atomic and Molecular Database
(LAMDA)\footnote{Available at http://home.strw.leidenuniv.nl/$\sim$moldata/}, which includes 22 and 24 energy levels for the ortho- and para-species, up to 554 and 294 cm$^{-1}$, respectively \citep{Schoier2005}. The collision rate
  coefficients  were also adopted from
  LAMDA, including temperatures ranging from 15 to 300 K.  The
optical depth was calculated for the \ammm~line for a density range of
10$^2-10^8$ cm$^{-3}$, a temperature range of 20--300\,K, and a
fractional abundance of \amm~of 10$^{-7}$. The results are plotted in
Fig.\ \ref{model}, showing a well defined peak of the optical depth around a density of 10$^5$\,cm$^{-3}$ with an increasing peak value for
warmer temperatures.

\section{Discussion}\label{discussion}

The fact that \meth~and \ammm~trace the same clumps of gas allow
estimates of the density and temperature in localized regions. If the
regions are within the positional error bars of $\gamma$-ray emission,
these estimates may provide interesting information of the conditions
of gas that may be linked with cosmic ray acceleration. Here we
discuss the density and positions of the shocked regions with respect to the SNRs.

\subsection{Density and Temperature}\label{density}

\citet{Mcewen2014} estimated the conditions conducive for \meth~36~GHz
and 44~GHz masers in SNRs. Their results show a comparably brighter
36~GHz maser in higher density regimes ($n>10^5$ cm$^{-3}$), and a
more dominant 44~GHz maser at lower densities ($n<10^5$~cm$^{-3}$) for
an optimal temperature $T<150$~K. The (3,3) inversion transition of
\amm~has a temperature of about 100 K above ground, and therefore the
emission distribution will emphasize excited gas. \ammm~maser emission
can occur via collisions with H$_2$ molecules \citep{Walmsley1983,
  Mangum1994, Zhang1995}.  Through statistical calculations it was found that \ammm~masers occur in gas with
temperatures around 200~K and densities on the order of $10^{3.5} -
10^7$ cm$^{-3}$ \citep[e.g.][]{Mauersberger1986,
  Mangum1994}. Consistent with previous calculations, the MOLPOP-CEP
estimates performed for our SNR environment (Sect.~\ref{modeling})
show optimal \amm~maser conditions at relatively high temperatures
($T>150$~K) and a relatively narrow density peak around $n\approx 10^5$ cm$^{-3}$ . At
those densities, there is, in fact, little difference in the inverted optical
depth between the 36~GHz and 44~GHz masers \citep{Mcewen2014}, perhaps
explaining the range of line ratios observed in the SNRs (Table
\ref{tbl-3}), where in some locations the 36~GHz maser is brighter, and the 44~GHz maser in others. That the line ratios of the 36~GHz and
44~GHz \meth~are on average unity further agrees with a relatively high
temperature in the clumps.  It thus
seems that the co-located clumps of \ammm~and \meth~masers trace gas
regions with $T>150$ K and $n\approx 10^5$\,cm$^{-3}$.

G1.4$-0.1$ is the exception, where both 36~GHz \meth~and \amm~emission
is profuse and 44~GHz is lacking. Here, the linewidths of the \amm~are
greater and the emission may be at least partly thermal. In this case,
the 36~GHz methanol might be tracing the densest clumps in the
interaction regions. A better understanding of the \amm~thermal versus
non-thermal emission here could be attained by also obtaining
observations of the NH$_3$(1,1) and (2,2) emission that may be
thermal, providing even stronger limits on the temperature and
density. Interferometric observations would also be helpful to tie
down the brightness temperature of the \ammm~lines.

\subsection{Spatial Trends}
The general picture of the spatial distribution of collisionally
pumped masers in SNRs was outlined in Sect. 1, where regions close to
the shock front harbor 36 GHz CH$_3$OH masers, and post-shock regions
harbor 1720 MHz OH masers. This depiction was based largely on
observations of Sgr A East, W28, and G1.4$-$0.1 \citep[e.g.,][]{Sjou2010,Nicholas2011,Pihl2011, Pihl2014}. With our
new data, this picture is further strengthened.

First, our comparison (Figs. $1-5$) of the NH$_3$, CH$_3$OH, and CO($1-0$) shows
a strong correlation in position and velocity, where both NH$_3$ and
CH$_3$OH are directly associated with the nearby MCs. As evidenced by the
CO maps, the extent of the MCs follows and is associated with radio
continuum contours. This is consistent with the presence of SNR/MC
interactions, providing the necessary excitation mechanism for both
NH$_3$ and CH$_3$OH. Second, the OH maser velocities listed in Table 1 are,
for most cases, offset in velocity from the CH$_3$OH and NH$_3$. Again, this
may imply that the OH is occuring further in the post-shock region,
resulting in a different velocity than the gas closer to the front.

An illustrating example is W44, which is interacting with a giant MC
(CO G34.875$-$0.625) with a V$_{LSR}$ around 48 km\,s$^{-1}$ \citep[e.g.,][]{Seta1998, Seta2004,dame2001, Yoshiike2013, Cardillo2014}. The brightest rim of this MC is overlapping with the brightest rim of the continuum emission (Fig. 3). The NH$_3$ and CH$_3$OH clumps coincide in position and velocity with the MC, but are offset from the OH detection.

Noteworthy details include the SNR candidate G5.7$-$0.0. The centroid
velocity of the NH$_3$ emission is around $-25$ km\,s$^{-1}$, almost 40 km\,s$^{-1}$
offset from the known OH maser at $+13$ km\,s$^{-1}$ \citep{Hewitt2009}, but instead matches well with the velocity of the CO emission detected
at the same position SW of the radio continuum (Fig. 2). To the NE,
closer to the OH maser, the CO gas instead has a higher, positive
velocity (3 to 23 km\,s$^{-1}$) agreeing with the OH velocity \citep{Liszt2009,dame2001}. This may be understood with the gas density differences
required for the excitation of the transitions.

\subsection{Association with $\gamma$-Ray Emission}
W44, W51C, and G5.7$-$0.0 are all detected in $\gamma$-rays \citep[e.g.,][]{Aharon2008, abdo2009, feinstein2009, Giuliani2011,Uchiyama2012}, which are thought to originate from the SNR/MC interaction.  In W44, for example, \citet{Uchiyama2012} report on using a pre-shock density of 200 cm$^{-1}$ and a post-shock density of $7\times10^3$ cm$^{-3}$, almost 15 times smaller than the post-shocked density close to the shock front we have derived.  If the $\gamma$-rays, just as the 36 GHz masers, are produced close to the shock front, our derived densities may provide additional information for the $\gamma$-ray emission estimated from neutral pion decay models.  

\section{Conclusions}\label{summary}
The combination of the presented GBT and VLA data have shown there is
a close correlation between \ammm~and \meth~36~GHz and 44~GHz maser
emission in SNRs interacting with MCs. That the \ammm~is truly maser
emission is not confirmed, but under this assumption the data is
consistent with clumps of gas of densities $n~\sim 10^5$ cm$^{-3}$ and
temperatures $T> 150$ K. 

\acknowledgments We would like to thank Joe Masters and Jim Braatz of
NRAO for their help with the GBTPIPELINE, Toney Minter for his help
with the GBT observations and calibration, Crystal Brogan for
providing the W51C and G5.7$-$0.0 continuum images and Betsy Mills for her discussion
about the ammonia observations. Finally, we would like to thank all
the staff scientists and operators at the GBT site for assisting with
the observations.  The National Radio Astronomy Observatory is a
facility of the National Science Foundation operated under cooperative
agreement by Associated Universities, Inc.  B.M. was supported by NASA
grant NNX10A055G.

\end{document}